**Parabolic-Dish Solar Concentrators of Film on Foam**

Sean A. Barton[1,2] and Ian L. Winger[1]

[1]Center for Materials Research and Technology (MARTECH), Florida State University, Tallahassee, Florida 32306-4351
[2]Department of Physics, Florida State University, Tallahassee, Florida 32306-4350

**Abstract**

Parabolic and spherical mirrors are constructed of aluminized PET polyester film on urethane foam. During construction, the chosen shape of the mirror is created by manipulating the elastic/plastic behavior of the film with air pressure. Foam is then applied to the film and, once hardened, air pressure is removed. At an *f*-number of 0.68, preliminary models have an optical angular spread of less than 0.25 degrees, a factor of 3.3 smaller than that for a perfectly spherical mirror. The possibility exists for creating large-lightweight mirrors with excellent shape and stiffness. These "film-on-foam" construction techniques may also be applicable to parabolic-trough solar concentrators but do not appear to be suitable for optical imaging applications because of irregularities in the film.

**Introduction**

Light-weight solar concentrators are of interest as the infrastructure required to support and rotate them is reduced compared to more massive concentrators. Light-weight concentrators must however still be sufficiently strong and stiff to maintain shape under wind loads. Parabolic troughs are of interest compared to parabolic dishes because they need to be rotated about only one axis to track the sun throughout the year with an attendant reduction in system complication. However, concentration ratios and working temperatures for troughs are correspondingly lower than those for dishes. High working temperatures are desirable to reduce the restriction of the Carnot limitation.

The theoretical concentration limit for troughs is about 273 suns. This corresponds to a limiting temperature of about 1500 K for an absorptive surface that is not wavelength selective when exposed to 1.0 suns (1000 W/m$^2$) of incident solar radiation (Stefan–Boltzmann law). With care, temperatures of 650 K are reached in practice [2]. This temperature corresponds to a maximum Carnot efficiently of 56%.

For parabolic dishes, the theoretical concentration limit is about 45,700 suns corresponding to a limiting temperature of about 5300 K (approximately the radiation temperature of the sun). At such high concentration ratios, heat must be transported away from the absorption site rapidly, otherwise high temperatures and attendant high conduction and radiation losses will occur. Advantageously, high concentration ratios do allow one to reach useful temperatures in conditions of heavy atmospheric filtration. To intentionally lower the concentration ratio, one can displace the absorbing surface away from the focal plane resulting in a sharply defined homogeneous disk of light. One should also note that the dish shape has excellent stiffness compared to the trough shape.  The drawback of needing to track the sun with two rotations still remains.  Currently all commercially successful solar-thermal power plants are of the trough type.

50  Solar concentrators have applications other than solar-thermal energy.  Parabolic and spherical mirrors can be used for production of artificial sunlight in the lab and for the creation of high-temperature solar furnaces.

## Methods and Techniques

55  The concentrating foamed dish is fabricated by applying foaming material onto a properly shaped reflective membrane.  For spherical and parabolic dishes, the reflective membrane is shaped by applying air-pressure to one side of the membrane.  Foaming material is then applied to the other side of the membrane in successive thin layers.  The foam is allowed to harden between layers.  When a sufficiently thick and stiff layer of foam has been created, the air

60  pressure is removed and the mirror is complete.

The current arrangement incorporates the use of a 0.001 in. thick aluminized PET polyester film as the reflective membrane (DuPont Mylar®), and a two-part liquid expanding urethane foam with an average cured density of 8 lb/ft$^3$ (U.S. Composites, Cat. No. FOAM-0804) as the

65  supporting foam structure for the membrane.

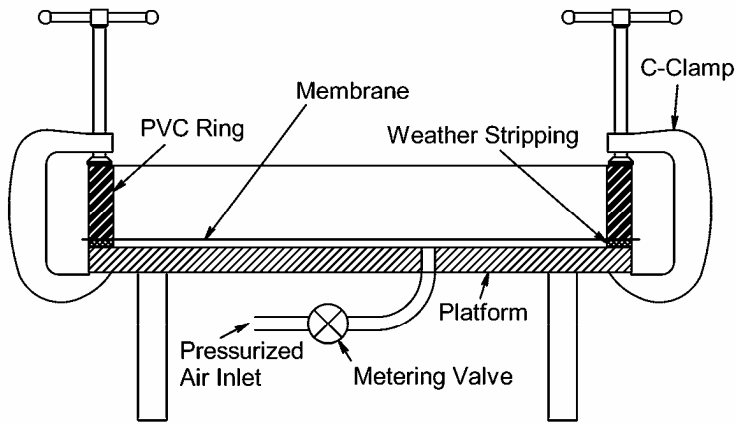

Figure 1, cross-section of the apparatus for making the dishes

70  A suitable pressurizing device is constructed to firmly hold the membrane along a circular edge while maintaining its planar integrity, as shown in Figs. 1 and 2.  The platform used for the current arrangement is made of a ½ in. thick 11 in. diameter aluminum plate with ½ in. wide weather stripping applied to the top edge.  The plate was elevated 3 in. by three legs to allow for the installation of a pressurizing line and metering valve underneath.  A 1.5 in. high PVC ring

75  with an outer diameter of 11 in. and internal diameter of 10 in. was used to clamp the edge of the membrane.  Eight standard "C" clamps were used around the periphery to supply the uniform clamping force needed to hold the film in place.  Air pressure is supplied by a diaphragm pump that is capable of at least 5 psi.

80    In assembling the system, the 0.001 in. thick reflective polyester film is cut to 12 in. diameter. This film is then centrally placed, reflective side down, on the platform. The PVC clamping ring is then positioned on top of the film, centered on the platform. The eight C-clamps are then arranged and tightened equally around the periphery to apply a sufficiently strong clamping force on the membrane to assure there will be no slippage of the film or air leakage.

85

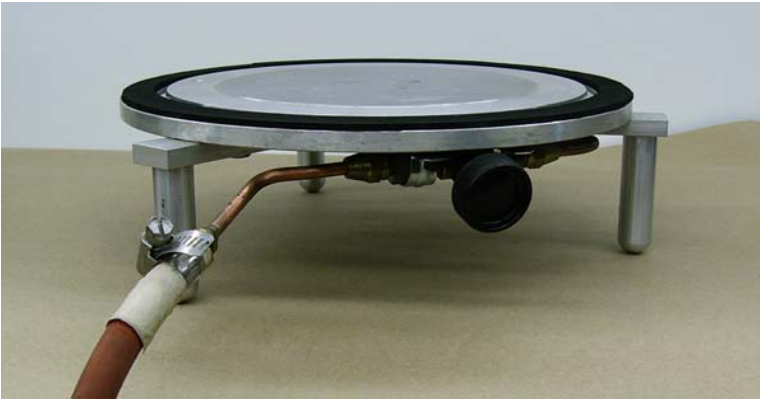

Figure 2, photograph of the baseplate of the apparatus for making the dishes

      Since the height of the PVC clamping ring is known to be 1.5 in., the top edge can be used as a
90    reference plane to measure the height of the inflated membrane. This was accomplished by placing a hard 12 in. rule on-edge across the pressure ring and zeroing a digital caliper at the distance to the uninflated film. Thus, after the film is inflated, the caliper will directly measure the actual height of the membrane (Fig. 3).

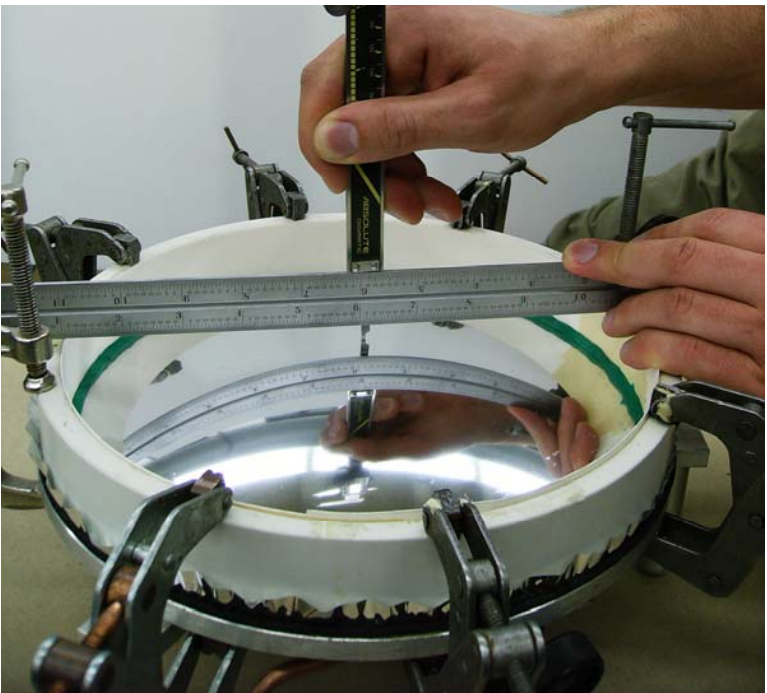

95    Figure 3, central displacement of inflation being measured

It must be noted that the process of properly layering the foam takes from 3 to 4 hours.  Thus, even very small leaks can seriously affect the final concentrator shape and its effectiveness.

100

The PVC ring also provides a barrier that prevents the liquid foam from escaping.  The urethane foam bonds aggressively to the PVC and will become permanently attached to it if they are allowed to contact directly.  It may be desirable to keep the PVC clamping ring attached to the concentrating foamed dish for mounting purposes.  If not, one can make a heavy paper ring to fit

105    on the inside of the PVC ring.  Applying a small bead of silicone caulking or modeling clay to the lower inside edge attaching the paper to the membrane will be sufficient to allow release of the foamed dish from the PVC ring after the foamed backing is completed.

Inflation of the membrane should be carried out by a procedure chosen to generate the desired

110    shape (sphere, parabola, etc.).  This procedure may involve several steps of inflation and/or deflation with gases and/or liquids.  In this example, we desire a parabola.  Experience suggests inflation with air to a central displacement of 1.064 in. followed by deflation to 0.907 in. (for a 10 in. diameter membrane).  When the proper procedure is followed the desired result can be verified by careful measurement on a "coordinate measuring machine" (CMM).

115

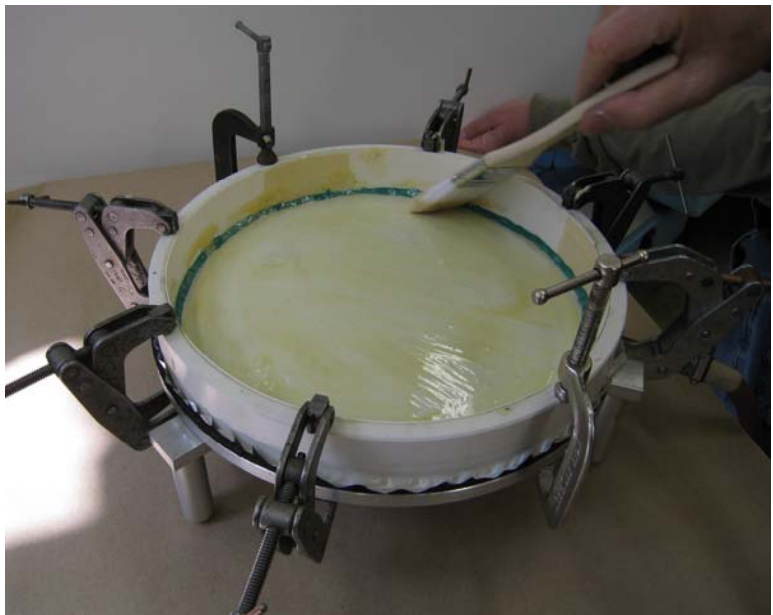

Figure 4, applying the first layer of foam

Pouring a large batch of uncured expanding foam completely over the membrane causes large

120    deformations of the concentrator because the foam tends to harden on the outside while the internal foam still wants to expand.  This creates local pressure irregularities on the inside of the foam that the tension in the film cannot overcome and thus the reflective surface becomes irregular instead of a smooth.  To overcome this problem, the uncured foam is brushed on the membrane in successive thin layers, allowing for the foam to completely cure between each

125    application (Fig. 4).  Six layers are usually sufficient to supply an adequate structural support for the film.

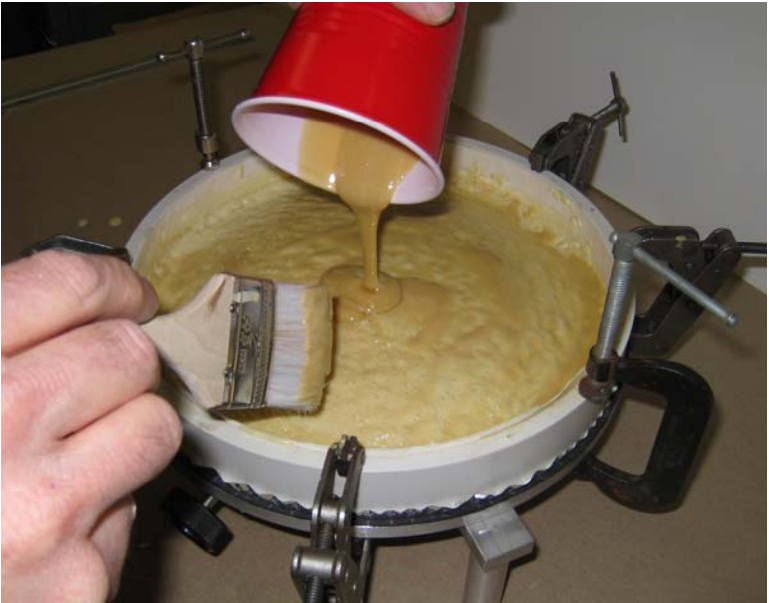

Figure 5, applying the fifth layer of foam

130

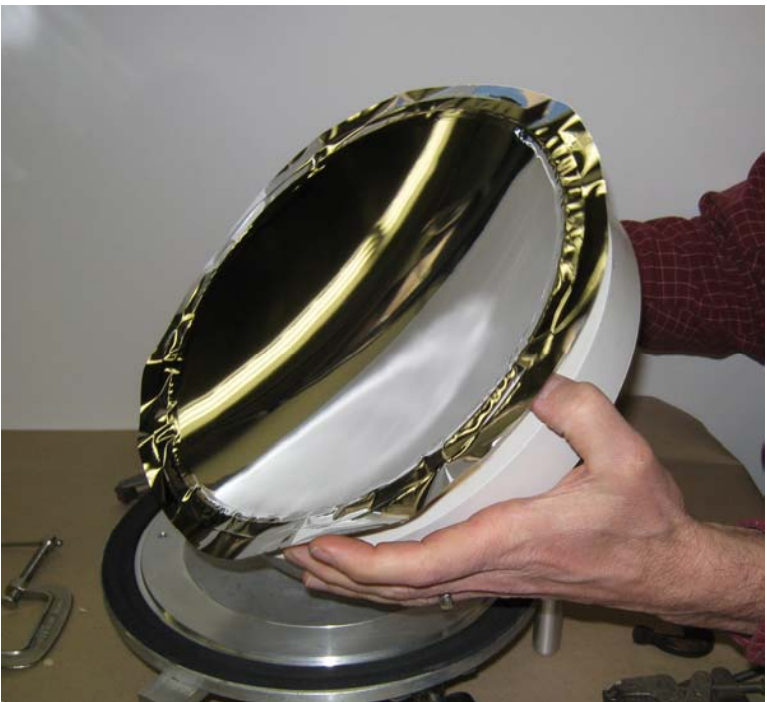

Figure 6, the finished dish held upright

135 The urethane foam used creates a very strong bond to the membrane.  To assure good adhesion to the film, it is critical that the first layer of uncured foam be painted on with many brush strokes.  This agitation mixes the uncured foam with any remaining trace oils or debris on the membrane that might compromise adhesion.  This first application of uncured foam should be very thin, but just enough to completely cover the film and insure a good hard crust of foam that will support the remaining layers.  The second layer should be painted on just as thin as the first 140   to ensure a strong supporting foundation for the subsequent layers. The remaining four layers can be applied progressively thicker than the previous ones (Fig. 5) until the foaming is complete.

When completed, the thickness of the foam at the center of the dish should be about 1/2 in., which can be measured from the PVC ring reference plane. After this thickness is achieved
145   (usually about 6 coats) the air pressure can be released and the clamps removed. Any foam that may have been inadvertently brushed on the PVC ring can now be scraped off and the finished foamed dish can be carefully removed from it (Fig. 6).

**Experimental Results**
150

Using an earlier and slightly larger apparatus than that discussed in section "Methods and Techniques" above, two membranes were characterized through several steps of inflation and deflation and then discarded without the application of foam. A first membrane of Mylar approximately 14 in. in diameter was inflated with a central displacement of about 1.36 in. A
155   profile of the shape was taken in the plane of the symmetry axis $\hat{z}$ and a diameter of the film $\hat{x}$ using two calipers. The resulting data were least-squares fit with four parameters to a quartic model (1).

$$z(x) = a_0 + a_2(x - x_0)^2 + a_4(x - x_0)^4 \tag{1}$$

160

The membrane was then deflated to a central displacement of about 0.81 in. and another profile was taken and these data were also fit to the model (1). The data and fits are shown in Fig. 7. Some datasets were taken only on half the diameter to speed the data taking process.

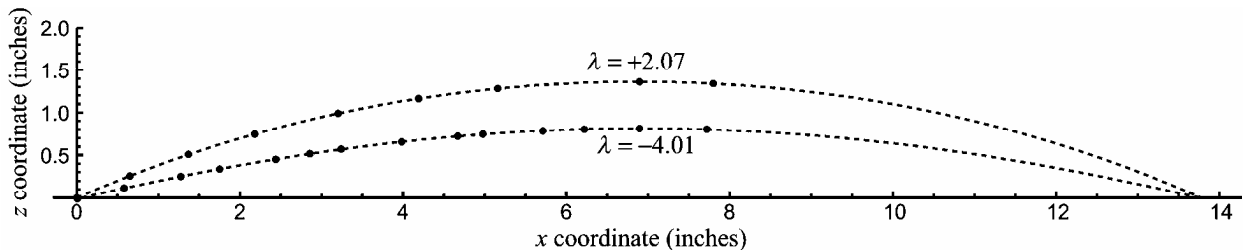

165

Figure 7, example inflation ($\lambda = +2.07$) and deflation ($\lambda = -4.01$) of a membrane showing super-spherical and super-hyperbolic behavior

We distinguish between spherical shapes and parabolic shapes by computing $\lambda \equiv a_4 / a_2^{3}$. As
170   explained below in the "Theory" section, $\lambda = 1$ describes a spherical shape and $\lambda = 0$ describes a parabolic shape. For the upper and lower curves of Fig. 7, $\lambda$ was found to be +2.07 and -4.01 respectively. This demonstrates that with an appropriate procedure, all shapes from super-spherical to super-hyperbolic (explained in the "Theory" section) are accessible experimentally.

175   A second membrane having the same diameter was inflated to a central displacement of 1.49 in. and then deflated in two steps to 1.32 in. and 1.27 in.. Profiles were taken at each step using a stylus on an XYZ stage. The profiles were fit to the same model (1). The data and fits are plotted in Fig. 8.

180  The $\lambda$'s were found to be +2.04, +0.31, and +0.08 respectively. A fourth profile was taken along a perpendicular diameter, immediately after the third profile. This $\lambda$ was found to be +0.86. This difference we attribute to the anisotropy of the film as described by the manufacturer [1]. The first three profiles were taken in the "machine direction" (MD) (see ref. [1]) and the fourth profile was taken in the perpendicular "transverse direction" (TD) (see ref. 185  [1]).

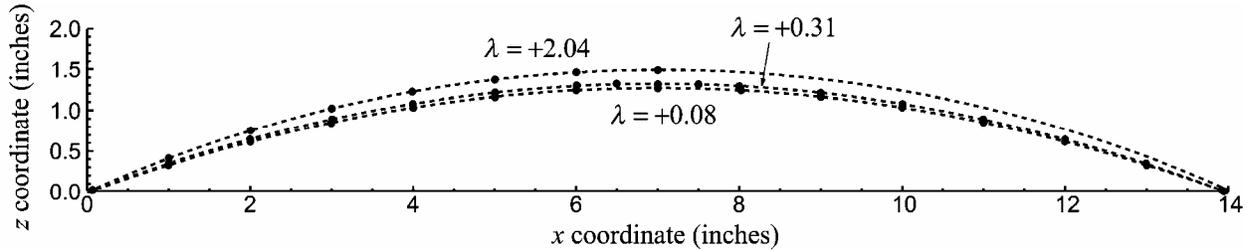

Figure 8, example inflation ($\lambda = +2.04$) and deflation in two steps ($\lambda = +0.31$ and $\lambda = +0.08$) of a membrane in attempt to find the parabolic shape ($\lambda_{\text{parabola}} \equiv 0$)

190

The raw data of the third profile (which is closest to the parabolic shape) are tabulated in Table 9. The data were taken in the listed order (not spatial order) to reduce the effect of a slow air leak. The model parameters for this data set were found to be $x_0 = +6.99949$ in., $a_0 = +1.26791$ in., 195  $a_2 = -2.6075 \cdot 10^{-2}$ in.$^{-1}$, and $a_4 = -1.33575 \cdot 10^{-6}$ in.$^{-3}$.

| $x$ (in.) | $z$ (in.) |
|---|---|
| 00.0645 | 0.0135 |
| 02.0020 | 0.6150 |
| 04.0011 | 1.0334 |
| 06.0014 | 1.2428 |
| 08.0219 | 1.2430 |
| 09.9998 | 1.0359 |
| 11.9976 | 0.6164 |
| 13.9477 | 0.0028 |
| 13.0023 | 0.3265 |
| 10.9995 | 0.8521 |
| 09.0022 | 1.1645 |
| 07.0017 | 1.2654 |
| 05.0032 | 1.1613 |
| 02.9960 | 0.8454 |
| 00.9966 | 0.3278 |

Table 9, an example numerical profile of an inflated membrane

The data of Table 9 were also fit to a sixth order model (2).

200  $$z(x) = a_0 + a_2(x - x_0)^2 + a_4(x - x_0)^4 + a_6(x - x_0)^6 \qquad (2)$$

The parameters were found to be $x_0 = +6.99949$ in., $a_0 = +1.26802$ in., $a_2 = -2.61185 \cdot 10^{-2}$ in.$^{-1}$, $a_4 = +1.1528 \cdot 10^{-6}$ in.$^{-3}$, and $a_6 = -3.45903 \cdot 10^{-8}$ in.$^{-5}$. A ray trace was then calculated on the sixth order model (2) for incoming parallel light. The location of each ray was found as it crossed the "focal plane" at $z = z_0 - 9.535$ in.. We define $\Delta x \equiv x - x_0$. In Fig. 10 we plot the $\Delta x$ for the ray's intersection with the focal plane verses the $\Delta x$ for the ray as it approaches from infinity.

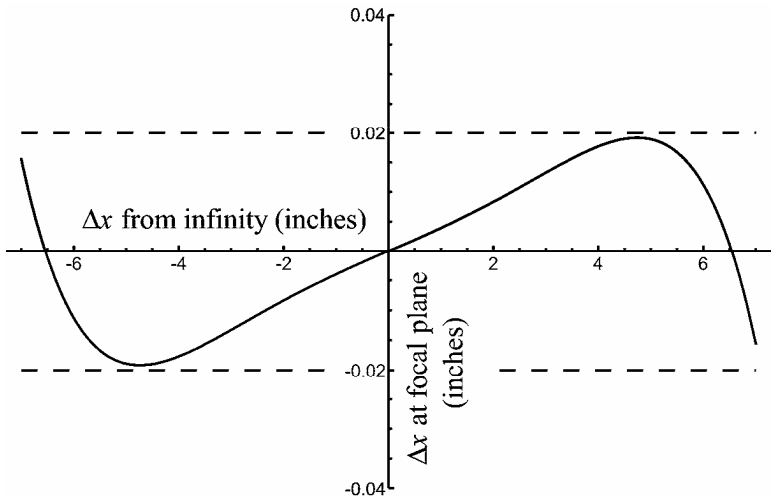

Figure 10, ray trace of the example profile

As one can see, all of the light is contained within a circle of diameter 0.040 in.. At this focal length, the additional spread of light corresponding to the angular diameter of the sun is about 0.088 in.. For a spherical mirror of the same focal length and aperture (radius of curvature 20.10 in.), the spread of the light resulting from spherical aberration would be approximately 0.133 in..

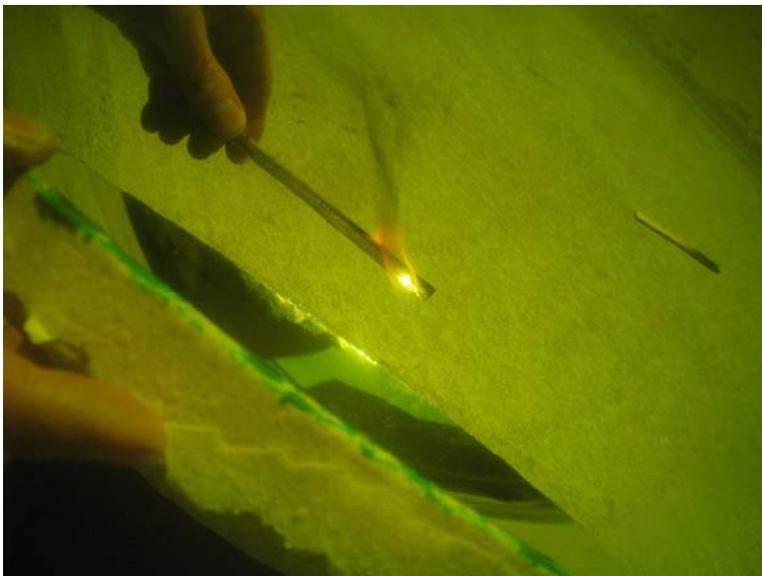

Figure 11, photograph of dish concentrating sunlight onto a wooden stick

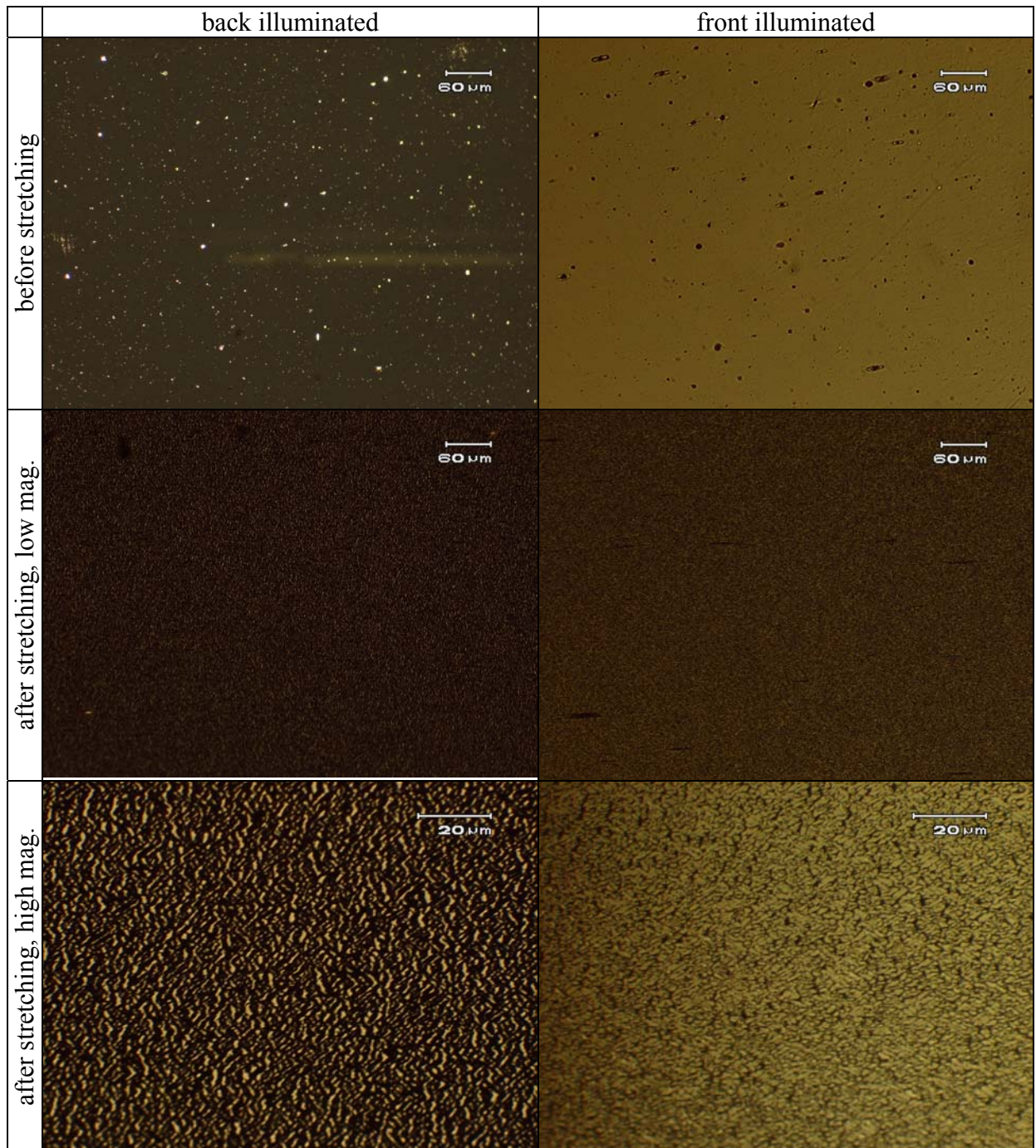

220 Figure 12, micrographs of aluminized Mylar before and after stretching about 200%: left column, back illuminated; right column, front illuminated; top row, before stretching; middle row, after stretching, low magnification; bottom row, after stretching, high magnification

Foam was not applied to the two membranes characterized above. However, foam was applied
225 to a different membrane of the same diameter for which exact profile information was not measured. This concentrator was then used to focus sunlight. A piece of wood held at the focal point caught fire within several seconds as seen in Fig. 11. The $\lambda$ for this membrane is not known.

230     It has been noticed that stretching the aluminized Mylar (aluminized PET polyester film) by a large factor causes the reflective surface to have a "foggy" appearance. Microscopic inspection (Fig. 12) indicates that the aluminum is breaking apart into "islands" of about 1 μm in size.

**Theory**

235

We consider an isotropic unstretched circular elastic membrane of radius unity that is fixed at its circumference. (The currently considered case maps trivially to membranes of arbitrary radius.) A differential pressure $P$ is applied across the membrane, stretching and distorting it. The differential pressure displaces the area elements of the film in the axial and radial directions, $\hat{z}$

240     and $\hat{r}$ respectively. We label a particular area element of the membrane by its original radius and azimuthal angle before inflation $s$ and $\phi$ respectively. Thus the shape of the membrane after inflation is described by three functions,

$$r \equiv r(s,\phi), \quad \theta \equiv \theta(s,\phi), \quad \text{and} \quad z \equiv z(s,\phi). \tag{3}$$

245

The system is cylindrically symmetric and thus we do not consider variations in the circumferential direction $\hat{\theta}$. Therefore

$$r(s,\phi) = r(s), \quad \theta(s,\phi) = \phi, \quad \text{and} \quad z(s,\phi) = z(s).$$

250

Before inflation $r(s) = s$ and $z(s) = \text{constant}$. The shape of the membrane after inflation is determined by minimizing the total potential energy. To calculate this, we define the following notational convenience.

255     $$\dot{r} \equiv r'(s), \quad \dot{z} \equiv z'(s), \quad \ddot{r} \equiv r''(s), \quad \text{and} \quad \ddot{z} \equiv z''(s) \tag{4}$$

We also define $\varepsilon_r$ and $\varepsilon_\theta$, the natural strain (or logarithmic strain) [3] in the radial and circumferential directions, respectively, for area elements of the membrane.

260     $$\varepsilon_r = \ln\sqrt{\dot{r}^2 + \dot{z}^2} \quad \text{and} \quad \varepsilon_\theta = \ln(r/s) \tag{5}$$

The total potential energy of the system $U$ can then be written as a functional of $r(s)$ and $z(s)$.

$$U = \int ds\, 2\pi s \left(\tfrac{1}{2}\alpha\varepsilon_r^2 + \tfrac{1}{2}\alpha\varepsilon_\theta^2 + \beta\varepsilon_r\varepsilon_\theta\right) + \int dV\, P \quad \text{where} \quad dV = 2\pi r z\, dr = 2\pi r \dot{r} z\, ds, \quad \text{or} \tag{6}$$

265

$$U = \int ds\, u \quad \text{where} \quad u(r,z,\dot{r},\dot{z}) = 2\pi s\left(\tfrac{1}{2}\alpha\varepsilon_r^2 + \tfrac{1}{2}\alpha\varepsilon_\theta^2 + \beta\varepsilon_r\varepsilon_\theta\right) + 2\pi r\dot{r}zP \tag{7}$$

Equation (7) is a sum of two terms, elastic energy and pressure-volume energy. Here $\alpha$ and $\beta$ are constants describing the elasticity of the membrane. In terms of Young's Modulus $E$ and 270 Poisson's Ratio $\nu$, $\alpha = E/(1-\nu^2)$ and $\beta = \nu E/(1-\nu^2)$. The state of the physical system is found by choosing $r(s)$ and $z(s)$ to minimize $U$ subject to the boundary conditions $r(0) = 0$, $r(1) = 1$, and $z(1) = \text{constant}$. We choose to represent $r$ and $z$ as power series expansions in $s$.

$$r = \sum_{n=0}^{\infty} r_n s^n \quad \text{and} \quad z = \sum_{n=0}^{\infty} z_n s^n \tag{8}$$

275

We set the boundary conditions by requiring $r_0 = 0$, $z_0 = 0$, $r_1 = \exp \varepsilon$, and $z_1 = 0$. In doing so we have enforced $r(0) = 0$, have set the axial coordinate system relative to the center of the membrane, and required smoothness at the center of the membrane. The central strain is $\varepsilon$ which depends on the applied pressure $P$. Our approach is to choose $\varepsilon$ and $P$ separately for

280 now and then later to choose the $P$ that satisfies the untreated boundary condition $r(1) = 1$. Calculus of variations provides the method by which we find $r(s)$ and $z(s)$.

$$\frac{\partial u}{\partial r} - \frac{d}{ds}\frac{\partial u}{\partial \dot{r}} = 0 \quad \text{gives}$$

$$\frac{\dot{r}}{\dot{r}^2 + \dot{z}^2}\left( \frac{\alpha s(\dot{r}\ddot{r} + \dot{z}\ddot{z})}{\dot{r}^2 + \dot{z}^2} + \frac{\beta s \dot{r}}{r} - \beta \right) + \frac{\alpha \varepsilon_r + \beta \varepsilon_\theta}{\dot{r}^2 + \dot{z}^2}\left( s\ddot{r} + \dot{r} - \frac{2s\dot{r}(\dot{r}\ddot{r} + \dot{z}\ddot{z})}{\dot{r}^2 + \dot{z}^2} \right) - \frac{s(\alpha \varepsilon_\theta + \beta \varepsilon_r)}{r} + r\dot{z}P = 0 \tag{9}$$

285

$$\frac{\partial u}{\partial z} - \frac{d}{ds}\frac{\partial u}{\partial \dot{z}} = 0 \quad \text{gives}$$

$$\frac{\dot{z}}{\dot{r}^2 + \dot{z}^2}\left( \frac{\alpha s(\dot{r}\ddot{r} + \dot{z}\ddot{z})}{\dot{r}^2 + \dot{z}^2} + \frac{\beta s \dot{r}}{r} - \beta \right) + \frac{\alpha \varepsilon_r + \beta \varepsilon_\theta}{\dot{r}^2 + \dot{z}^2}\left( s\ddot{z} + \dot{z} - \frac{2s\dot{z}(\dot{r}\ddot{r} + \dot{z}\ddot{z})}{\dot{r}^2 + \dot{z}^2} \right) - r\dot{r}P = 0 \tag{10}$$

Equations (9) and (10) represent a linear system in $\ddot{r}$ and $\ddot{z}$. We substitute into these equations

290 second order power series expansions for $r$ and $z$ and first order power series expansions for $\dot{r}$ and $\dot{z}$. Since $r_2$ and $z_2$ are still unknown, these power series expansions have unknown coefficients. This linear system is solved for $\ddot{r}$ and $\ddot{z}$ as functions of the unknown coefficients and these are then set equal to the $\ddot{r}$ and $\ddot{z}$ arrived at by taking second derivatives of $r$ and $z$ directly which also contain the unknown coefficients. This results in a pair of linear equations

295 relating the unknown coefficients to themselves. This linear system is solved to give the unknown coefficients $r_2$ and $z_2$. This process can now be repeated keeping one more term in every power series expansion. In this way $r_3$ and $z_3$ are found. One can repeat this procedure as many times as desired to solve for an arbitrary number of the $r_n$ and $z_n$. We find $r_{2n} = 0$ and $z_{2n+1} = 0$ for all integers $n$. Furthermore we find

300

$$z_2 = \frac{P \exp(4\varepsilon)}{4(\alpha + \beta)\varepsilon}, \quad r_3 = \frac{P^2 \exp(7\varepsilon)(3\alpha - \beta - 2\alpha\varepsilon - 2\beta\varepsilon)}{64(\alpha + \beta)^2 \varepsilon^2 (\alpha\varepsilon + \beta\varepsilon - \alpha)}, \quad \text{and}$$

$$z_4 = \frac{P^3 \exp(10\varepsilon)(\alpha-\beta)(6\varepsilon-1)}{512(\alpha+\beta)^3 \varepsilon^4 (\alpha\varepsilon+\beta\varepsilon-\alpha)}. \tag{11}$$

305    With $r(s)$ and $z(s)$ now known to 4th order, we can now find $z(r)$ to 4th order.

$$z(r) = \sum_{n=0}^{\infty} a_n r^n, \quad a_0 = 0, \quad a_{2n+1} = 0,$$

$$a_2 = \frac{P \exp(2\varepsilon)}{4(\alpha+\beta)\varepsilon}, \quad \text{and} \quad a_4 = \frac{P^3 \exp(6\varepsilon)\left(-\alpha+\beta-6\alpha\varepsilon-2\beta\varepsilon+8\alpha\varepsilon^2+8\beta\varepsilon^2\right)}{512(\alpha+\beta)^3 \varepsilon^4 (-\alpha+\alpha\varepsilon+\beta\varepsilon)}. \tag{12}$$

310

As the distinction between parabola and sphere is of interest in this work, we define the quantity $\lambda \equiv a_4/a_2{}^3$. It can be shown, by making power series expansions of the conic sections, that $\lambda$ values of +1, 0, and -1 correspond to spheres, parabolas, and hyperbolas respectively. It should be noted that the $\lambda$'s calculated earlier from the parameters of (1) are representative of the entire dish. The currently calculated $\lambda$'s are representative only of the center point of the dish. In $\lambda$ 315    we recognize $\beta/\alpha$ as Poisson's Ratio $\nu$. Making this substitution, we finally find that

$$\lambda = \frac{(1-\nu)+(6+2\nu)\varepsilon-(8+8\nu)\varepsilon^2}{8\varepsilon-(1+\nu)\varepsilon^2}. \tag{13}$$

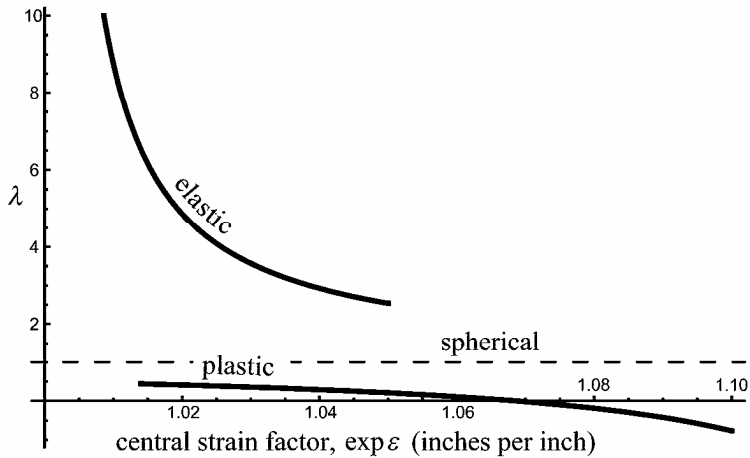

320

Figure 13, $\lambda$ as a function of central strain factor, $\exp\varepsilon$

Notice that $\lambda$ does not depend on $P$ or $\alpha$ and $\beta$ separately. It can be shown that variations in $P$ or variations in $\alpha$ and $\beta$ that do not affect $\nu$ only change the overall linear dimension of the solution (i.e. $z(r) \to Cz(r/C)$ where $C$ is some overall scale factor). These changes do not 325    affect the shape of the solution (sphere, parabola, etc.). Varying $P$ in this way makes satisfying the boundary condition $r(1) = 1$ trivial. In the case of Mylar where $\nu = 0.38$ [1], we can plot this $\lambda$ as a function of the central strain $\varepsilon$ (Fig. 13, curve labeled "elastic").

330    The elastic limit for Mylar at room temperature is approximately 1.4% [1]. Thus at $\varepsilon > 1.014$ plasticity plays some role in the shape of the membrane and this calculation of $\lambda$ becomes inaccurate. However we can see that the shapes achievable up to the elastic limit are "super-spherical". To facilitate discussion, we now define the following terms.

335    $\begin{cases} \lambda > +1 & \text{"super-spherical"} \\ \lambda = +1 & \text{"spherical"} \\ \lambda < +1 & \text{"sub-spherical"} \\ \lambda > 0 & \text{"sphere-like"} \\ \lambda = 0 & \text{"parabolic"} \\ \lambda < 0 & \text{"hyperbola-like"} \\ \lambda > -1 & \text{"sub-hyperbolic"} \\ \lambda = -1 & \text{"hyperbolic"} \\ \lambda < -1 & \text{"super-hyperbolic"} \end{cases}$

Notice that a particular value of $\lambda$ may be described by several of these.

We now wish to model the membrane in the plastic regime. We can model this approximately
340    by making the assumption of "proportional loading". Under this assumption, the strain tensor during inflation in each area element of the membrane is a monotonically increasing power of some generating matrix that is constant for that particular area element. Strictly, this assumption is incorrect. It is correct at the center of the membrane and at the edge of the membrane but not in the annulus in between. A fully numerical simulation is needed to properly treat plasticity.
345    But for now we make the assumption of proportional loading to find analytic results that are of broad relevance. We also make the assumption of "perfect" plasticity (the absence of strain hardening). In contrast to elastic energy, which is proportional to the square of the strain, the proportional-loading plastic energy pseudo-potential is proportional to the absolute value of the strain. Thus we replace (7) with (14).

350
$$u(r, z, \dot{r}, \dot{z}) = 2\pi s \sqrt{\alpha \varepsilon_t^2 + \alpha \varepsilon_\theta^2 + 2\beta \varepsilon_t \varepsilon_\theta} + 2\pi r \dot{r} z P \tag{14}$$

Now $\alpha$ and $\beta$ are constants that describe the plasticity of the membrane. Following the same procedure as before we find

355
$$z_2 = \frac{P \exp(4\varepsilon)}{2\sqrt{2}\sqrt{\alpha + \beta}}, \quad r_3 = -\frac{P^2 \exp(7\varepsilon)(\alpha - \beta - \alpha\varepsilon - \beta\varepsilon)}{8(\alpha + \beta)(\alpha - \beta - 2\alpha\varepsilon - 2\beta\varepsilon)},$$

$$z_4 = -\frac{3P^3 \exp(10\varepsilon)(\alpha - \beta)}{32\sqrt{2}(\alpha + \beta)^{3/2}(\alpha - \beta - 2\alpha\varepsilon - 2\beta\varepsilon)}, \quad a_2 = \frac{P\exp(2\varepsilon)}{2\sqrt{2}\sqrt{\alpha + \beta}},$$

360

$$a_4 = \frac{P^3\exp(6\varepsilon)(\alpha - \beta - 4\alpha\varepsilon - 4\beta\varepsilon)}{32\sqrt{2}(\alpha + \beta)^{3/2}(\alpha - \beta - 2\alpha\varepsilon - 2\beta\varepsilon)}, \quad \text{and} \quad \lambda = \frac{(1-\nu) - (4 + 4\nu)\varepsilon}{(2 - 2\nu) - (4 + 4\nu)\varepsilon}. \tag{15}$$

For Mylar in the plastic regime $\nu \approx 0.58$ [1]. Again we plot $\lambda$ as a function of the central strain $\varepsilon$ (Fig. 13, curve labeled "plastic"). We notice that in the plastic regime that $\lambda$ is much lower (sub-spherical). We further notice that when $\varepsilon = \varepsilon_{\text{parabola}} \equiv (1-\nu)/(4+4\nu)$, $\lambda = 0$ (parabolic).

365 When $\nu = 0.58$, $\exp\varepsilon_{\text{parabola}} = 1.06871$. Thus, it appears that the sphere is accessible by stopping the inflation in the range of the elastic limit and that the parabola is accessible by inflating somewhat further.

To further investigate non-proportional loading and the possibility of plastic inflation followed
370 by elastic deflation, a fully numerical model is needed. By varying the amount of inflation and deflation, it may be possible to vanish both $a_4$ and $a_6$ in (2). To date, this numerical model has been developed only to the point that it has verified the analytic model of elastic inflation.

**Conclusion**

375

This is a work in progress. It is clear that fully numerical models of the membrane are desired and effort is currently directed at obtaining these. Also experimental efforts at fabricating larger dishes appear to be attractive and useful. The current limitation is determined by the available
380 widths of aluminized polyester film (a maximum of 50 in.). It is interesting to explore methods by which several 50-in. widths of film might be seamed together in a way that does not interfere with the homogeneous plasticity of the composite membrane. One possibility is to create the membrane of two layers of film oriented perpendicular to each other. This may also eliminate the slight anisotropy of the membrane related to the method of manufacturing the film [1]. It is
385 also interesting to attempt to use this technique to produce parabolic troughs. Though simple inflation of a long rectangular membrane will always result in a cylindrical surface, it may be possible to introduce an additional inflatable bladder (or perhaps several coaxially) under the center-line of the rectangle to approximate the parabolic shape.

**Acknowledgements**

390

The authors would like to acknowledge Steve Bellenot for generously devoting hours of his time to help us discover the fault in our code, Anter El-Azab for helping us to understand plasticity, Peter Höflich for providing computing resources, Frank Flaherty and Sanford Safron for useful discussion, David Van Winkle for unwavering support, and Pedro Schlottman and Efstratios
395 Manousakis for discussion of computational techniques. The authors would also like to give a major acknowledgement to Dan Baxter and the staff of the Physics Machine Shop (Randall, Terri, Jim, Greg, Rick, and Zane) for generous use of their resources. Without them this project would not be possible.